\documentclass[sigconf]{acmart}
\usepackage{natbib}
\usepackage{graphicx}
\usepackage{textcomp}
\usepackage{float}
\usepackage[hang,flushmargin]{footmisc}
\usepackage{import}
\usepackage{color}

\usepackage{soul}
\usepackage{pifont}
\usepackage{enumitem}
\usepackage{multirow}
\usepackage{blindtext}
\usepackage{dirtytalk} 

\usepackage{tabularx} 
\usepackage{multirow}
\usepackage{listings} 
\usepackage{todonotes}
\lstdefinestyle{python}{ 
    xleftmargin=6.0ex,
    xrightmargin=2.0ex,
    numbers=left,
    frame=single
}
\definecolor{gray10}{gray}{.9}
\definecolor{arsenic}{rgb}{0.23, 0.27, 0.29}
\usepackage{color}

\RequirePackage{fontawesome}

\definecolor{gray50}{gray}{.5}
\definecolor{gray40}{gray}{.6}
\definecolor{gray30}{gray}{.7}
\definecolor{gray20}{gray}{.8}
\definecolor{gray10}{gray}{.9}
\definecolor{gray05}{gray}{.95}

\newlength\Linewidth
\def\findlength{\setlength\Linewidth\linewidth
  \addtolength\Linewidth{-4\fboxrule}
  \addtolength\Linewidth{-3\fboxsep}
}
\newenvironment{examplebox}{\par\begingroup
  \setlength{\fboxsep}{5pt}\findlength
  \setbox0=\vbox\bgroup\noindent
  \hsize=0.95\linewidth
  \begin{minipage}{0.95\linewidth}\normalsize}
  {\end{minipage}\egroup
  \textcolor{gray20}{\fboxsep1.5pt\fbox
    {\fboxsep5pt\colorbox{gray05}{\normalcolor\box0}}}
  \endgroup\par\noindent
  \normalcolor\ignorespacesafterend}

\begin{document}

\title{Ignoring Time Dependence in Software Engineering Data. \\ A Mistake}

\author{Mikel Robredo}
\affiliation{%
  \institution{University of Oulu}
  \city{Oulu}
  \country{Finland}
}
\email{mikel.robredomanero@oulu.fi}

\author{Nyyti Saarim\"{a}ki}
\affiliation{%
  \institution{Tampere University}
  \city{Tampere}
  \country{Finland}
}
\email{nyyti.saarimaki@tuni.fi}

\author{Rafael Pe\~naloza }
\affiliation{%
  \institution{University of Milano-Bicocca}
  \city{Milano}
  \country{Italy}
}
\email{rafael.penaloza@unimib.it}

\author{Valentina Lenarduzzi}
\affiliation{%
  \institution{University of Oulu}
  \city{Oulu}
  \country{Finland}
}
\email{valentina.lenarduzzi@oulu.fi}

\renewcommand{\shortauthors}{Robredo, et al.}

\begin{abstract}
Researchers often delve into the connections between different factors derived from the historical data of software projects. For example, scholars have devoted their endeavors to the exploration of associations among these factors. However, a significant portion of these studies has failed to consider the limitations posed by the temporal interdependencies among these variables and the potential risks associated with the use of statistical methods ill-suited for analyzing data with temporal connections. Our goal is to highlight the consequences of neglecting time dependence during data analysis in current research. We pinpointed out certain potential problems that arise when disregarding the temporal aspect in the data, and support our argument with both theoretical and real examples.
\end{abstract}




\maketitle

\section{Introduction}
\label{sec:Introduction}
In Empirical Software Engineering (ESE), particularly within the field of Mining Software Repositories (MSR), researchers frequently explore relationships between various variables derived from the historical data of software projects. For instance, scholars have dedicated their efforts to examining correlations among variables such as code smells, design smells, and architectural smells~\cite{PalombaTSE2018, ArcelliFontana2019, Sharma2020}, scrutinizing their temporal evolution~\cite{Olbrich2009}, and assessing the influence of different software attributes~\cite{Sjoberg2013, Palomba2018, lenarduzzi2020some}.

Nevertheless, a substantial portion of these investigations has neglected to address the constraints imposed by the temporal interdependence of these variables and the potential vulnerabilities associated with the application of statistical techniques ill-suited for the analysis of temporally linked data. For example, introducing a code smell in a software commit heavily relies on the state of the codebase preceding that commit. Yet, most studies have disregarded this critical facet, primarily due to the absence of well-defined statistical methodologies for handling such temporal dependencies.

From prior publications within the Mining Software Repository field, we can pinpoint three primary concerns:
\begin{enumerate}
\item Discarding the temporal nature of the commits
\item Assumption of independent data
\item The proportional size of projects, where big projects overwhelm small ones. 
\end{enumerate}
In order to further investigate this existing issue, we aim at visualizing the impact that ignoring time dependence on data analysis has on the current research. We identify some potential issues when ignoring the time dependence in the data and present theoretical and real examples. 

\textbf{Paper Structure:} Section~\ref{sec:Issue} visualizes the issues that current analysis techniques face. Section~\ref{sec:Examples} presents theoretical and real-life examples to better comprehend the existence of this paper's background problem. Section~\ref{sec:Approach} depicts the defined roadmap to design the future analysis guidelines and Section~\ref{sec:Conclusion} presents the conclusions of this paper.
\section{Current Issues}
\label{sec:Issue}

\begin{figure}[h!]
    \centering
    \includegraphics[width=1\columnwidth]{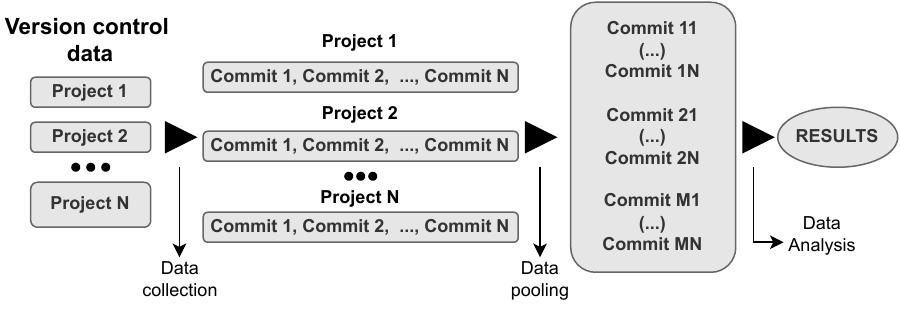}
    \caption{Current data collection and analysis approach in MSR studies (N: Nº of projects, M: Nº of commits)}
    \label{fig:current-approach}
\end{figure}

The possibility of obtaining data through the mining of software projects' version control systems has opened a wide source of data for current research. In particular, MSR studies utilize features such as commits, refactorings, and releases located in the source code of projects to study their quality~\cite{Olbrich2009, Sjoberg2013, Palomba2018, lenarduzzi2020some, Olbrich2010, AmpatzoglouTSE2015, khomh2009exploratory, amanatidis2017relation, baldassarre2020diffuseness}. Frequently in these and other studies of similar nature, researchers tend to collect the version control data of the selected projects during the data collection process, and subsequently pool the considered metrics all together into the same dataset. Then, the considered data analysis is performed on the dataset to obtain a global result. Figure~\ref{fig:current-approach} presents graphically the described procedure.

As briefly mentioned in the introduction, there are three main concerns about the current research approach in MSR:
\begin{itemize}
    \item \textbf{Discarding the temporal nature of commits:} Commits are temporally ordered within the version control of projects. Each commit is dependent on what existed before, and therefore there exists a time dependency between commits. For instance, an existing bug in the current commit may be the result of mistakes in the code developed in previous commits, or colloquially speaking ``A mistake in the code today may lead into a bug tomorrow.'' This potential relationship vanishes when all commits from different projects are pooled together. 
    
    \item \textbf{Assumption of independent data:} Most of the mentioned studies in the current literature perform the commonly used statistical methods such as Mann-Whitney, Wilcoxon, or ANOVA in their data analyses~\cite{Wohlin2000}. Among the multiple assumptions these methods require is the independence of observations inside groups. While pooling the projects' commits together, it is most likely that time-dependent commits from the same projects may end up in the same group. Therefore, as the assumption of independence among groups is not fulfilled, the results from the analysis cannot be entirely correct.
    
    \item \textbf{Dimensional difference among projects:} To ensure the generalizability of the results, multiple studies consider including projects of different dimensions. However, often all projects are pooled together regardless of the number of commits the different projects have. Hence, when the data pooling is performed, the weight of bigger projects exceeds that of smaller projects, removing representativity in the study from the second group.
\end{itemize}


Ignoring these factors in the nature of the data can lead to negative consequences and errors in the subsequent results. Hence, the researcher must carefully choose which statistical methods to use in the data analysis based on the assumptions fulfilled by the distribution of the data. If the researchers fail at this stage, some of the potential negative effects are the following:

\begin{itemize}
    \item \textbf{Considering wrong assumptions:} Most statistical methods currently in use rely on different sets of assumptions that the data points must respect. If time dependence is not considered among the assumptions of the chosen statistical method to analyse time-dependent data, the obtained results can lead into misleading interpretations and poor model performance.

    \item \textbf{The threat of losing temporal relationships:} Time dependent data often exhibit patterns between consecutive data-points. Losing the temporal relationship among observations could yield wrong results which do not identify valuable insights or anomalies, e.g. concluding that mistakes in the code and bugs are not related.

    \item \textbf{Misleading results:} In an era where good results can mask possible misinterpretations of the data, the mistake of not considering time-dependent methods for time-dependent data can cause the resulting analysis to hide the true nature of the data. That is, results relying on wrong methods can yield misleading interpretations, e.g. concluding that more mistakes in the code can lead to fewer bugs.
\end{itemize}

In essence, the consequences of a bad administration of the code are not obvious at an initial stage after the mistake has been made. Time-dependent methods raise as a powerful approach to analyse the spread of the created mistakes across the time. Similarly, assessing bad practices in the code based on the prevention of future bugs may not have instant improvements, especially for big projects where multiple professionals are involved. Still, this practice will make a positive impact in the long run. 
\section{Examples in Software Engineering}
\label{sec:Examples}
We now present two examples that depict the described current issues in the ESE research field. We first introduce a theoretical example with the aim of better conceptualizing the possible consequences of ignoring time dependence when analysing time-dependent data. In the second example we describe a real application of the current data analysis approach in MSR studies by observing the work by \citeauthor{lenarduzzi2020some}~\cite{lenarduzzi2020some}.

\subsection{Theoretical example} \label{sec:theoExample}

\begin{figure}[tbh]
    \centering
    \rotatebox{270}{\includegraphics[width=0.70\columnwidth]{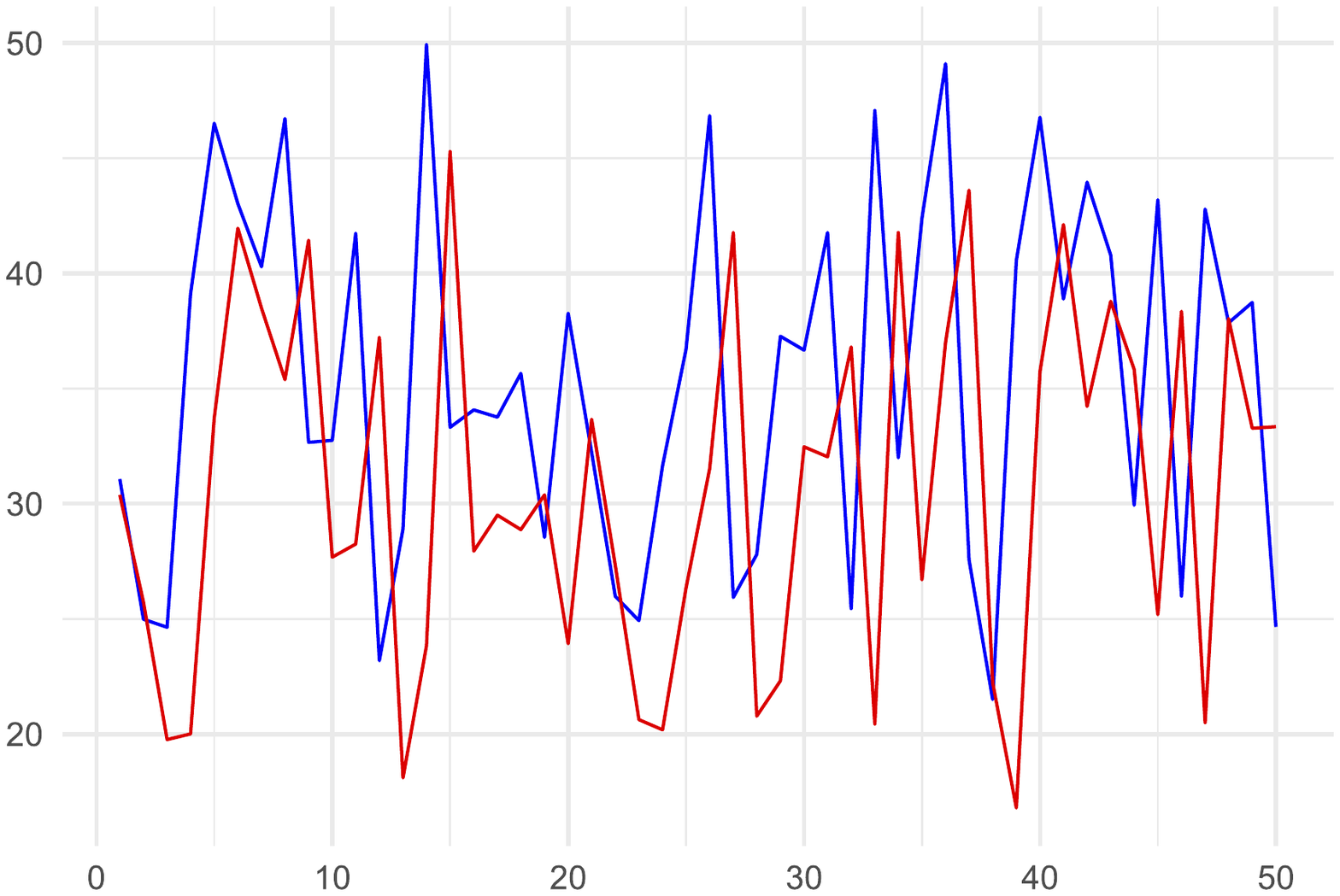}}
    \caption{Temporal evolution of the example variables.}
    \label{fig:example_vars}
\end{figure}

Let us consider a theoretical example showing the importance of considering the assumption of time dependence when analysing time-dependent data. In this example there is a \textit{dependent} variable (\textcolor{red}{red}) which at time \textbf{$t + 1$} is highly correlated to the \textit{independent} variable (\textcolor{blue}{blue}) at time \textbf{$t$}. Figure~\ref{fig:example_vars} depicts the first 50 observed points of the temporal relationship within the data.

\begin{figure}[b]
    \centering
    \rotatebox{270}{\includegraphics[width=0.70\columnwidth]{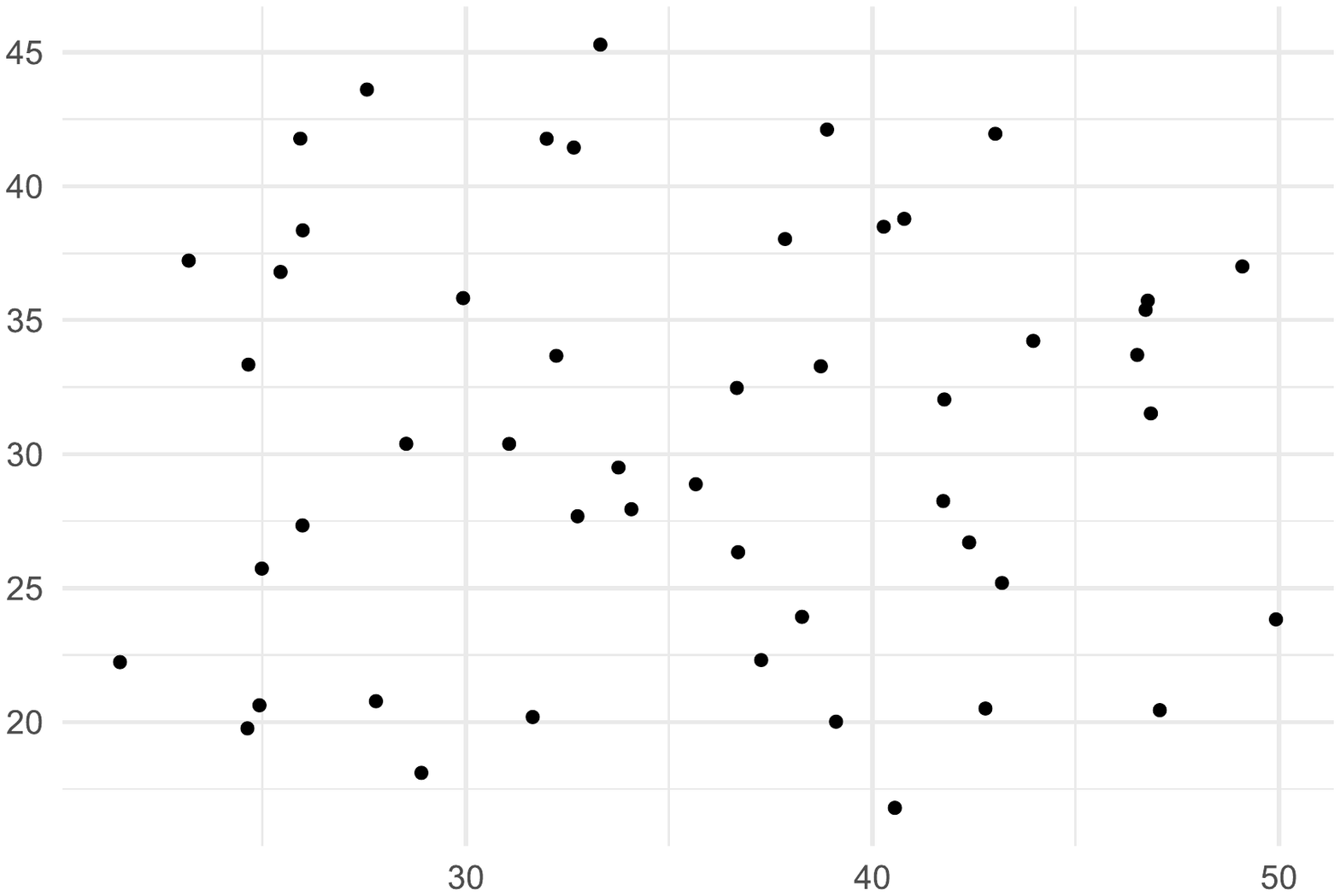}}
    \caption{Correlation plot of the example variables considering the same time point.}
    \label{fig:uncorrelation}
\end{figure}

If we were not conscious about the temporal relationship of the data, the omission of the time dependence between the described variables could lead us to consider erroneous assumptions. In fact, an exploratory correlation plot of such scenario could be depicted as in Figure~\ref{fig:uncorrelation}. At a first glance, this perspective of the example variables could lead the researcher to discard the existing time dependence.

We need to dedicate importance to this stage in the analysis of the two variables if the assumption of time dependence was removed. An erroneous conclusion about the correlation between the variables would lead to a misinterpretation, for instance, a nonexistent correlation between the variables. 

\begin{figure}[h!]
    \centering
    \rotatebox{270}{\includegraphics[width=0.70\columnwidth]{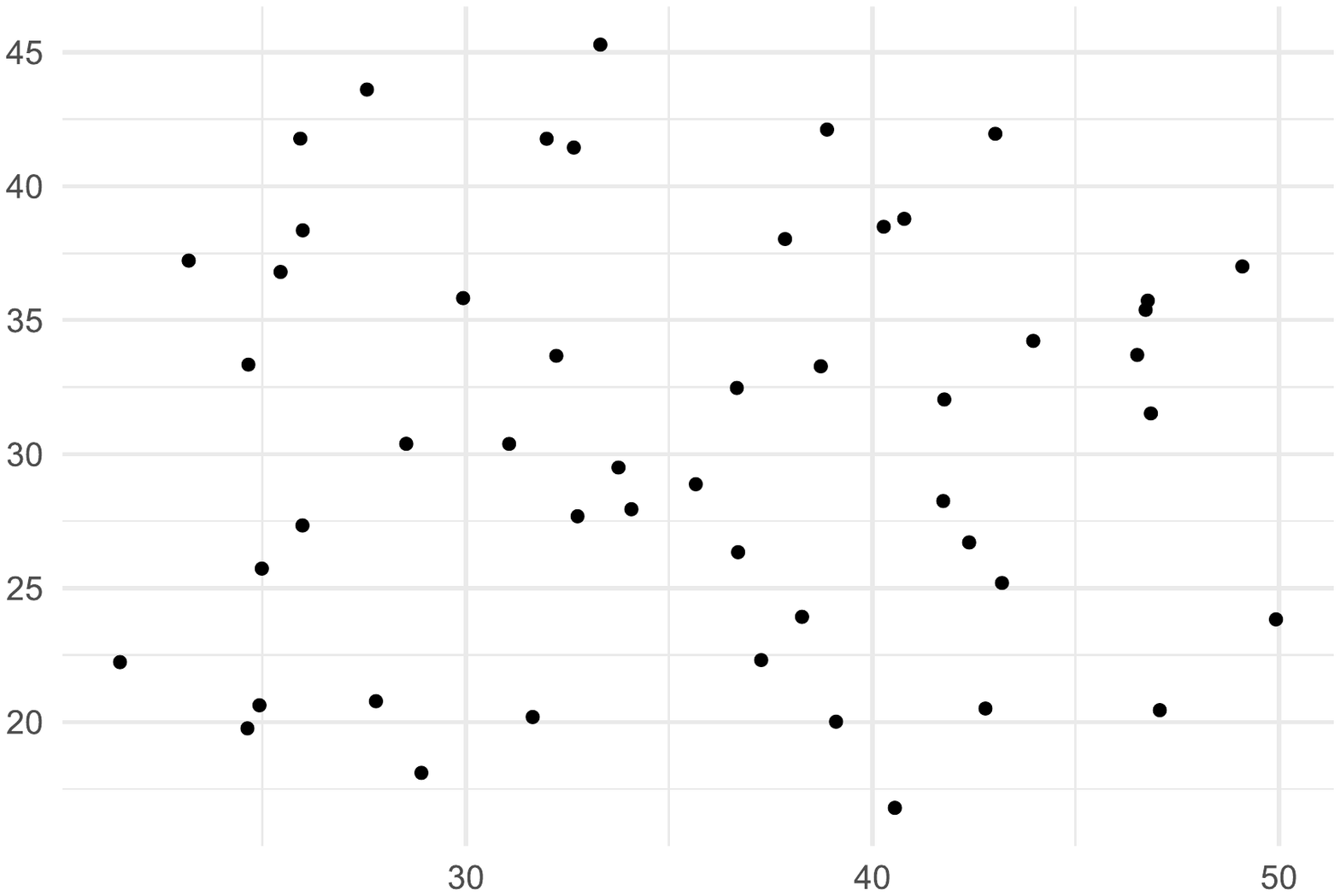}}
    \caption{Correlation plot of the example variables considering a lag of one time point between them.}
    \label{fig:correlation}
\end{figure}

If the researcher considers the assumption of an existing time dependence among the dependent variable and the independent variable, the exploratory analysis of the variables could explain that the dependent variable is the same as the independent variable in the previous point given a sort of small variables. Once this assumption is considered, the correlation plot of the variables displayed in Figure~\ref{fig:example_vars} would depict an existing temporal relationship, similar to the one displayed in Figure~\ref{fig:correlation}.

\subsection{Real example} \label{sec:realExample}
Let us now consider a real case application of current data collection and data analysis approach in MSR studies. The chosen example is the study performed by \citeauthor{lenarduzzi2020some}~\cite{lenarduzzi2020some}, which investigates the diffuseness of SonarQube issues on 33 Java projects from the Apache Software Foundation, and assesses their impact on code changes and fault-proneness, considering also their different types and severities. To accomplish that, they investigated the differences between classes that are not affected by Sonar issues (\textbf{clean}) and classes affected by at least one Sonar issue (\textbf{dirty}). Thus, their study compares the change-proneness and fault-proneness of the classes within these two groups.

\begin{figure}[h!]
    \centering
    \includegraphics[width=1\columnwidth]{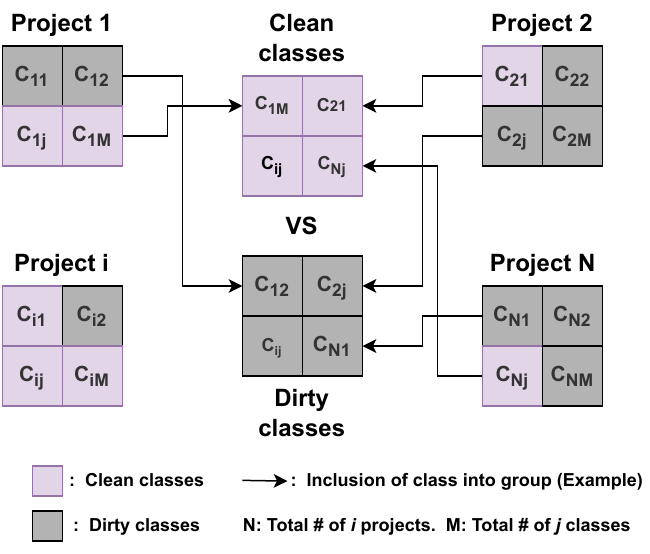}
    \caption{Graphical representation of the class groups built by pooling all together.}
    \label{fig:class_pooling}
\end{figure}

The study collected commits from each of the mentioned 33 projects once every 180 days (average time between releases in the analyzed projects). Subsequently, they calculated the change-proneness and fault-proneness for each of the existing classes from the current commit to the incoming commit within all projects.  
To compare the distribution of the two groups of classes, they used two widely used non-parametric statistical tests. First, they used the \textit{Mann-Whitney} test to compare the statistical significance of each of the group distributions. This statistic is used to test whether the distributions of two groups are statistically different, i.e. whether the distributions of observations for both groups of variable of interest have the same shape or a different shape. Furthermore, the method chooses random samples from both groups in order to perform the comparison, given the assumption of independence among observations in each group.
Secondly, they used the \textit{Cliff's Delta} effect size test to investigate the magnitude of the size difference between the two groups. In either case, these tests were used with the preliminary assumption of independence among the group observations.

The described study concluded that classes affected by Sonar issues may be more change- and fault-prone that non affected classes. However, the differences found between the clean and affected classes resulted to be small. 

Taking a closer look into the described data analysis, the \textit{Mann-Whitney} statistic assumes independence among the observations in each group, which in this case are classes in the dirty and clean groups. For instance, assume that a given project has multiple clean and dirty Java classes and consider the scenario in which classes are connected between each other within the same project, so that changes in one of the classes may have an impact in the other classes in the short or long run. If two or more connected classes from the same project were located inside the same group, the assumption of independence between observations of the same group would be erroneous, therefore the results from the performed analysis would lead into misleading interpretations. 

Moreover, within the considered 33 Java projects large projects overwhelmed small ones, which was reflected in the number of commits collected for each of the projects. This fact can create a dimensional gap in impact of the considered projects in the study, therefore the weight of the biggest projects may exceed that of the smaller projects. This makes the results of this study lack from generalizability since the representativity of all projects is not complete.

\section{The Roadmap}
\label{sec:Approach}
Our proposed roadmap includes the following steps: 

\begin{enumerate}
    \item \textbf{Investigation of the state-of-the-art data analysis approaches adopted in the ESE domain.} We will conduct a literature review (e.g. Systematic Literature Review~\cite{KITCHENHAM} or Systematic Mapping Study~\cite{PETERSEN}) of the existing data analysis approaches and methods adopted in Software Engineering to analyse the collected data. This step will allow us to have a complete and exhaustive knowledge of the current situation in order to identify biases and issues. 
    \item \textbf{Identification of data analysis issues in the Empirical Software Engineering studies.} Based on the results of the previous step, we will identify potential issues in the current data analysis approaches that can help us in the new approach definition.
    \item \textbf{The new approach conceptualization.} Based on the previous two steps, we will define the first draft in a iterative way for a new data analysis approach addressing the identified issues. 
    \item \textbf{Comparison of the new approach with the traditional ones adopted by Empirical Software Engineering studies}. To demonstrate the accuracy of our new model, we will perform a large scale of empirical studies to compare it with the current approaches.
    \item \textbf{Internal Validation }(Application in Empirical Software Engineering domain). Our aim is to evaluate the proposed methodology by selecting a set of published Empirical Software Engineering studies. We will replicate these studies with our methodology and compare the results. The outcome of the analysis will be discussed with experts in Empirical Software Engineering to evaluate the validity of the methodology. 
    \item \textbf{Guidelines for data analysis definition.} We will elaborate a protocol on how to apply this approach.
\end{enumerate}

\section{Conclusion}
\label{sec:Conclusion}

In this paper, we conceptualize the issues in the current research approach in MSR studies and discuss the negative effects that this approach can have in the data analysis results. In concrete, we insist on the vital assumption of time dependence in commit data, and therefore advocate for its consideration in the current research.

In order to defend our position, we define a a roadmap that includes the necessary steps to redefine the current guidelines of the data analysis methodology in Software Engineering.



\bibliographystyle{ACM-Reference-Format}
\bibliography{sample-bibliography.bib}

\end{document}